\documentclass[prd,aps,twocolumn,superscriptaddress,preprintnumbers,
showpacs,tightenlines,floatfix,nofootinbib,amssymb]{revtex4}

\pdfoutput=1 

\usepackage{epsfig}

\usepackage[english]{babel}
\usepackage{amsmath}
\usepackage{mathtools}
\usepackage{relsize}
\usepackage{epsfig}
\usepackage{setspace}
\usepackage{subfigure}
\usepackage{graphics}
\usepackage{graphics}
\usepackage{newlfont}
\begin{document}
%
%
\title{Search for fingerprints of disoriented chiral condensates in cosmic ray showers}

\author{R. M. de Almeida}\email{rma@ifi.unicamp.br}
\affiliation{Instituto de F\'isica Gleb Wataghin, Universidade Estadual de Campinas, 
Caixa Postal 6165, Campinas, SP 13083-970, Brazil}
\affiliation{Instituto de F\'isica, Universidade Federal do Rio de Janeiro, 
Caixa Postal 68528, Rio de Janeiro, RJ 21941-972, Brazil}
\author{J. R. T. de Mello Neto}\email{jtmn@if.ufrj.br}
\affiliation{Instituto de F\'isica, Universidade Federal do Rio de Janeiro, 
Caixa Postal 68528, Rio de Janeiro, RJ 21941-972, Brazil}
\author{E. S. Fraga}\email{fraga@if.ufrj.br}
\affiliation{Instituto de F\'isica, Universidade Federal do Rio de Janeiro, 
Caixa Postal 68528, Rio de Janeiro, RJ 21941-972, Brazil}
\author{E. M. Santos}\email{emoura@if.ufrj.br}
\affiliation{Instituto de F\'isica, Universidade Federal do Rio de Janeiro, 
Caixa Postal 68528, Rio de Janeiro, RJ 21941-972, Brazil}

\begin{abstract}
Although the generation of disoriented chiral condensates (DCCs), where the order 
parameter for chiral symmetry breaking is misaligned with respect to the vacuum 
direction in isospin state, is quite natural in the theory of strong interactions, they 
have so far eluded experiments in accelerators and cosmic rays. If DCCs are formed 
in high-energy nuclear collisions, the relevant outcome are very large event-by-event 
fluctuations in the neutral-to-charged pion fraction. In this note we search for fingerprints 
of DCC formation in observables of ultra-high energy cosmic ray showers. We present 
simulation results for the depth of the maximum ($X_{max}$) and number of muons on 
the ground, evaluating their sensitivity to the neutral-to-charged pion fraction asymmetry 
produced in the primary interaction.
\end{abstract}


\maketitle

\section{Introduction}

Almost forty years ago exotic, apparently hybrid and unexpected events, dubbed Centauros, 
were observed in cosmic ray (CR) experiments in emulsion chambers in Chacaltaya by 
Lattes and collaborators \cite{Lattes:1971qw}. Those events were very different from what is 
commonly observed in CRs, exhibiting a large number of hadrons and a small number of electrons 
and gammas, which suggests the presence of very few rapid-gamma-decaying hadrons. So, a 
possible imbalance in the number of neutral to charged pions could be envisaged. The nature 
and reality of Centauro events started a long debate, that includes the reexamination of the 
original emulsion chamber plates, and is still not 
resolved \cite{Augusto:1999vz,GladyszDziadus:2001cq,Kopenkin:2003pi,Engel:2005gf}.

Nevertheless, Centauro events were certainly an experimental motivation for the development 
of the theory of disoriented chiral condensates (DCCs) that started in the early 
1990s \cite{Anselm:1991pi,Bjorken:1991sg,Blaizot:1992at,Rajagopal:1992qz}. 
For a detailed review, see Ref. \cite{Mohanty:2005mv}. Assuming that a given nuclear 
system could be heated above the critical (crossover) transition region for chiral symmetry 
restoration, i.e. for temperatures of the order of $180-190~$MeV \cite{Bazavov:2009zn}, then 
quenched to low temperatures, the chiral condensate initially melted to zero could grow in any 
direction in isospin space. Besides the vacuum (stable) direction, it could build up as in a 
metastable, misaligned pseudo-vacuum state, and later decay to the true, chirally broken 
vacuum. The fact that DCCs could be formed in high-energy heavy ion collisions stimulated 
several theoretical advances and experimental searches \cite{Mohanty:2005mv}. Most 
likely the temperatures achieved in current heavy ion experiments are high enough to 
produce an approximately chirally symmetric quark-gluon plasma, and the following rapid 
expansion can cool the system back to the vacuum \cite{QM2009}, so that the dynamics of 
chiral symmetry restoration and breakdown can be described in a quench 
scenario \cite{Rajagopal:1992qz}, so that the evolution of the order parameter can be much 
affected by an explosive behavior that naturally leads to large fluctuations and 
inhomogeneities \cite{quarks-chiral,Mishustin:1998yc,Scavenius:1999zc,paech,Aguiar:2003pp}.

Since, by assumption, the order parameter for chiral symmetry breaking, i.e. the chiral 
condensate, is misaligned with respect to the vacuum direction (associated with the 
$\sigma$-direction in effective models for strong interactions) in a DCC, this would be a 
natural candidate to explain the excessive production of hadrons unaccompanied by 
electrons and photons, suggesting the suppression of neutral pions with respect to charged 
pions.

Regardless of the outcome of the debate on the nature of Centauro events, DCC formation 
seems to be a quite natural phenomenon in the theory of strong interactions. However, given 
its symmetric nature (in isospin space), it should be washed out by standard event averaging 
methods. So far, there has been no evidence from colliders or CR experiments. Motivated by 
the possibility of attaining much higher statistics in current ultra-high energy cosmic ray (UHECR) 
experiments than in the past, so that an event-by-event analysis for very high-energy collisions 
can in principle be performed, we consider possible signatures of DCC production in CR air 
showers. If DCCs are formed in high-energy nuclear collisions in the atmosphere, the relevant 
outcome from the primary collision are very large event-by-event fluctuations in the 
neutral-to-charged pion fraction, and this could affect the nature of the subsequent atmospheric 
shower. Very preliminary, yet encouraging results were presented in Ref. \cite{Viana:2007zz}.

In this paper we search for fingerprints of DCC formation in two different observables of UHECR 
showers. We present simulation results for the depth of the maximum ($X_{max}$) and number 
of muons on the ground, evaluating their sensitivity to the neutral-to-charged pion fraction asymmetry 
produced in the primary interaction. To model the effect from the presence of a DCC, we simply 
modify the neutral-to-charged pion fraction, assuming that the system follows the same 
kinematics, as will be detailed below. Although this is certainly a very crude description of the 
dynamics of the primary collision, we believe it captures the essential features that have 
to be tested in order to verify the feasibility of detecting DCCs in UHECR showers.

This paper is organized as follows. In Section II we briefly review some characteristics of 
DCCs, especially the Baked-Alaska scenario and the inverse square root distribution of 
the neutral pion fraction. In Section III the method for the simulation is presented. We 
use CORSIKA \cite{CORSIKA}, a program for detailed simulation of extensive air showers 
initiated by high-energy cosmic ray particles. In Section IV we show and discuss our results. 
Section V contains our conclusions.

\section{DCC features and the neutral pion fraction distribution}

It is widely believed that for high enough energy densities, achieved e.g. by increasing 
dramatically the temperature, strong interactions becomes approximately chiral (it would 
be an exact symmetry only if current quarks were strictly massless), so that the chiral 
condensate, which is the order parameter for that transition, essentially vanishes. On the 
other hand, for low temperatures the chiral condensate acquires a non-vanishing value 
and breaks spontaneously the chiral symmetry of (massless) QCD \cite{Yagi:2005yb}.

In a given model, one can construct an effective potential for the chiral condensate degrees 
of freedom and study the mechanism of chiral symmetry restoration and breakdown. If we 
restrict our analysis to two flavors of light quarks, up and down, that can be easily accomplished 
by the linear sigma model coupled to quarks \cite{GellMann:1960np}. In that case, the 
effective degrees of freedom are pions, $\pi^{a}$, and the sigma, $\sigma$. In the high-temperature 
limit all field expectation values vanish, whereas in the vacuum one has $\langle\pi^{a}\rangle =0$ 
and $\langle\sigma\rangle =f_{\pi}$, where $f_{\pi}$ is the pion decay constant.

The physical picture we have in mind is a very high-energy heavy ion collision that will 
create a hot quark-gluon plasma where chiral symmetry is approximately restored. As the 
plasma is quenched to low temperatures by expansion, the system will evolve to the 
vacuum emitting a large number of pions. However, the evolution can proceed along 
many different paths in chiral space before it finally reaches the true vacuum, i.e. it can 
``roll'' into different directions, and the ratio of neutral to charged pions produced depends 
strongly on the chosen metastable state in each event. In other words, the misalignment 
of the vacuum is reflected in the distribution of produced pions, generating a coherent state 
and an anomaly in the ratio of charged to neutral pions. This effect will be, of course, 
washed out by event averages.


A more intuitive physical picture for the formation of DCCs, the {\it Baked-Alaska} scenario, 
was proposed by Bjorken and 
collaborators \cite{Bjorken:1991sg, Bjorken:1993cz, AmelinoCamelia:1997in}. 
Consider a high-multiplicity collision at high energy. Before hadronization, most of the energy 
released at the collision point is carried away by the primary partons at nearly the speed of 
light. This hot and thin expanding shell isolates the relatively cold interior from the 
outer vacuum. As the state evolves, the interior cools down and what was before a global 
minimum becomes a local maximum. The symmetry is spontaneously broken and one of the 
pseudo-vacua should be chosen, with a slight preference for the true vacuum. However, if 
the lifetime of the shell is short enough, the quark condensate in the interior might be rotated 
from its ordinary direction, since it costs relatively little energy, 
$\epsilon \sim f^{2}_{\pi}  m^{2}_{\pi} \approx 20$ MeV/fm$^3$ \cite{Mohanty:2005mv}. 
When the hot shell hadronizes, the reorientation induced by the external contact is reflected 
in the produced pions, which will be coherent and present fluctuations in an event-by-event 
analysis. 

Assuming that the correlated region is large enough to be described semiclassically, one 
can use the linear sigma model with explicit symmetry breaking for the description of the 
dynamics \cite{AmelinoCamelia:1997in}. Quantitatively, one can represent the shell of this 
fireball as the source of these excitations of pions: 
\begin{equation}
(\Box +m^{2}_{\pi}) \pi^{a}(x) = J^{a}(x) \,,
\end{equation}
where $a$ is the isospin index. A spherically expanding shell is represented by
\begin{equation}
J^{a}(x) = J^{a}(t) \Theta(t-t_0) \delta (t-r) \,,
\end{equation}
with initial radius $r_0=t_0$, where $t_0$ is the time where the expansion starts. 
After the hadronization, the currents $J^{a}(t)$ vanish and the fields $\pi^{a}$ decay 
towards the vacuum into freely propagating pions. So, the pion emission is characterized 
by the state
\begin{equation}
| \vec{J} \rangle   = {\cal N} \exp \left(\sum^{3}_{a=1} \int \frac{d^{3}\vec{k}}{(2\pi)^3} 
J^{a}(\vec{k}) c^{\dagger}(\vec{k}) \right) |0 \rangle \,,
\end{equation}
where ${\cal N}$ is a normalization factor, and the sum is over isospin directions $a=1,2,3$. 
The creation operator of a pion with momentum $\vec{k}$ and isospin component $a$ is 
$c_{a}^{\dagger}(\vec{k})$ and $J_{a}(\vec{k})$ is the 4-dimensional Fourier transform of the 
source $J^{a}(x)$ at $k^{0}=\sqrt{\vec{k}^{2}+m_{\pi}^{2}}$. 

The number of pions follows a Poisson distribution with average
\begin{equation}
\overline{n}^{a}(\vec{k}) = \langle \vec{J} | c^{\dagger}_{a}(\vec{k}) c_{a}(\vec{k}) | \vec{J} \rangle = 
\left.| J^{a}(\vec{k})|\right.^{2}    \, ,
\end{equation}
and the number of pions produced per unit phase space is approximately given 
by \cite{Mohanty:2005mv}
\begin{equation}
\frac{dN_{a}^{(J)}}{d^{3}k} \approx \frac{\left.| J^{a}(\vec{k})|\right.^{2}}{(2\pi)^3} \, .
\end{equation} 
%
 
 
Statistically, one expects that the magnitude and the chiral orientation of the source $J^{a}(\vec{k})$ 
will fluctuate event by event for each mode. Let us assume that the vacuum orientation is tilted into 
one of the pion directions, i.e.:
\begin{equation}
\langle \sigma \rangle = f_{\pi} \cos \theta   \ \ \ ,   
\ \ \langle \vec{\pi} \rangle = f_{\pi} \sin \theta \ \hat{u} \, , 
\end{equation}
so that all relevant modes of chiral condensate point in the same direction $\hat{u}$ in 
isospin space:
\begin{equation}
\vec{J}_{DCC}(\vec{k}) = J(\vec{k}) \hat{u} \, .
\end{equation}
In generic models of production the neutral pion fraction, defined as
\begin{equation}
f\equiv \frac{N_{\pi^0}}{N_{\pi^0}+N_{\pi^+}+N_{\pi^-}} \, ,
\end{equation}
is a binomial distribution with average $\overline{f}=1/3$. In this way, in the limit of large 
numbers $N_{\pi^{a}}$, the probability of all pions being charged is very small. However, 
if pions are the product of a DCC decay, this probability is not negligible. In fact, if there is 
no privileged isospin direction, the vector $\hat{u}$ can point in any direction within the unit 
sphere. Then:
\begin{equation}
f_{DCC}=\cos^{2} \theta \, ,
\end{equation}
where $\theta$ is the angle between the unit vector $\hat{u}$ and the $\pi^0$ direction. So, 
one finds the following well-known distribution for the neutral pion 
fraction \cite{Bjorken:1991sg,Blaizot:1992at}:
\begin{equation}
dP \approx d(\cos \theta) = \frac{1}{2\cos \theta} d(\cos^{2} \theta) = \frac{df}{2\sqrt{f}} \, .
\label{sqrt}
\end{equation}
%
The probability of less than 10$\%$ of pions be neutral, for instance, is $\sim 30\%$.  

\section{DCC simulations}

The conditions of a high temperature initial state followed by a rapid cooling stage are 
both possible to happen in heavy ion collisions at the very high energies like those 
accessible at RHIC and to be reached at the LHC, as well as in UHECR collisions in the 
top of the atmosphere. The large aperture of a detector like that of the Pierre Auger 
Observatory \cite{EA} (combining shower sampling at ground level and longitudinal shower profile 
reconstruction) has been providing high quality data and unprecedented statistics in the 
field \cite{xmax_measure, Abraham:2010mj, Abraham:2010yv}. Therefore, even though the formation 
of a DCC is probably rare, we believe it is worth studying the implications of such events to the 
physics of showers generated by UHECR.

The neutral pion fraction distribution, Eq. (\ref{sqrt}), is at the basis of our strategy to search 
for DCC fingerprints in UHECR showers. Therefore, any investigation to measure the impact caused 
by the presence of a DCC should assess the sensitivity of a given observable with respect to the 
neutral pion fraction produced in the primary interaction. This fraction determines the initial 
distribution of particles between the electromagnetic and hadronic components of the showers. In this 
paper, we consider two observables which are usually measured by UHECR detectors: the slant depth in the 
atmosphere (defined as the integral of the atmosphere density along the shower axis\footnote{For brevity, 
from now on we should refer to the slant depth as simply depth. However, the reader should be aware that 
in the literature the term depth may be used to refer to the integral of the atmosphere density along 
the vertical and not along the shower axis.} $\int\rho dx$ and expressed in units of g/cm$^2$) at which 
the shower reaches its 
maximum development, $X_{max}$, and the number of muons on the ground $N_{\mu}$. It is known that the 
parameter $X_{max}$ is affected by the first interaction cross-section and its associated multiplicity 
and inelasticity \cite{Engel:2005gf}. 

If DCCs really exist, the conditions for them to be produced should include not only high 
energy densities, but the regions where such densities are achieved should not be small as well, 
since DCCs are considered ``macroscopic'' space-time regions where the chiral parameter is not oriented 
in the same direction as the vacuum. With those requirements in mind, we have chosen to work with 
Fe initiated showers at $10^{19}$ eV. Then, central collisions are privileged over peripheral ones 
by selecting events with a large number of participating nucleons ($N_{part}>40$) which, in turn, 
should lead to a higher multiplicity in the first interaction.

For all simulations presented in this work, we have used CORSIKA 6.617 \cite{CORSIKA} with the 
interaction models Sibyll 2.1 \cite{Ahn:2009wx} and GHEISHA 2002d \cite{gheisha}, for high and low 
energy processes, respectively. The DCC-like shower simulation chain is as follows: Large $N_{part}$ 
collisions are selected and their first secondaries (mostly pions and kaons) separated; after, some 
of the neutral pions in this sample are converted into charged ones; the resulting particle list is 
then inserted back into CORSIKA to proceed with usual cascade development through the atmosphere. Such 
a procedure was performed for several $\pi^{0}$ fractions and 2 different zenith angles. The first 
interaction slant depth ($X_{0}$) distribution for DCC-like showers will therefore be the same as for 
normal Fe initiated showers in central collisions. This is a valid assumption, since the DCC formation 
process takes place during a subsequent cooling stage of the initial hot plasma, with the first interaction 
cross section being the same as in a Fe-nucleus collision. We believe that even though the simulation 
approach adopted here is simplified, the essential features of the process are being taken into account. 

For comparison, proton initiated showers as well as normal Fe initiated ones were also generated. For the 
normal Fe case, we have produced both a sample rich in central collisions and another sample with all the 
centralities. For each shower we extract the value of $X_{max}$ and the number of muons on the ground.

\section{Discussion of results}

\begin{figure}[!h]
	\begin{center}
		\scalebox{0.9}{\includegraphics[scale=0.48]{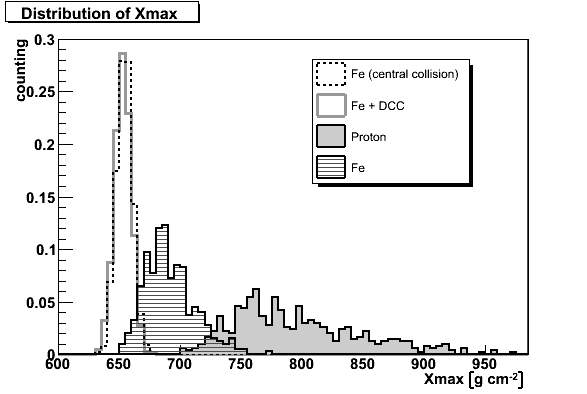}}
		\caption{Normalized $X_{max}$ distributions for vertical showers initiated by Fe (central collision), 
		Fe+DCC, Fe and proton with $E=10^{19}$ eV.}
\label{xmax}
	\end{center}
\end{figure}

From now on we shall identify our DCC-like Fe initiated showers by Fe+DCC. Fig. \ref{xmax} shows the $X_{max}$ 
distribution of a vertical shower ($\theta=0$) corresponding to an extremely asymmetric situation ($f=0$), 
that is, where all the initially 
produced $\pi^{0}$'s are converted into $\pi^{+}/\pi^{-}$. For the distribution of Eq. (\ref{sqrt}), sharply 
peaked at $f=0$, the probability for less than 1\% of $\pi^{0}$'s being produced is 10\%. Four types of showers: 
normal Fe (central collision: Fe-Central), Fe (all centralities), Fe+DCC and proton initiated are shown. On can 
clearly see that both samples generated from a central collision have smaller than average $X_{max}$, since the higher the multiplicity in the first interaction the faster is the subsequent cascade 
development in the atmosphere. And there is essencially no difference between the Fe-Central and the Fe-DCC 
samples in terms of $X_{max}$, which might indicate that the early stage where the $\pi^{0}$ population is depleted 
together with the higher interaction probability due to the large multiplicity allow for a complete recovery of 
these particles in the subsequent interactions. Nonetheless, it is clear that one should look at low $X_{max}$ 
events when searching for DCCs signatures, and this property is independent of the initial $\pi^{0}$ fraction.

Another feature which can be appreciated in Fig. \ref{xmax} is the smaller $X_{max}$ fluctuations for 
Fe-DCC and Fe-Central as compared to those of proton and Fe with all centralites. This is to be expected, since the fluctuations in $X_{max}$ have basically two components: the the ones in the first 
interaction slant depth $X_{0}$ (which, in turn, depends on the interaction cross-section) and those 
introduced by the cascade growth up to the maximum. The latter depends on the initial shower size 
(the multiplicity), in average being smaller for high multiplicity events, i.e. those found in the central 
collisions.

\begin{figure}[!h]
	\begin{center}
		\scalebox{0.9}{\includegraphics[scale=0.48]{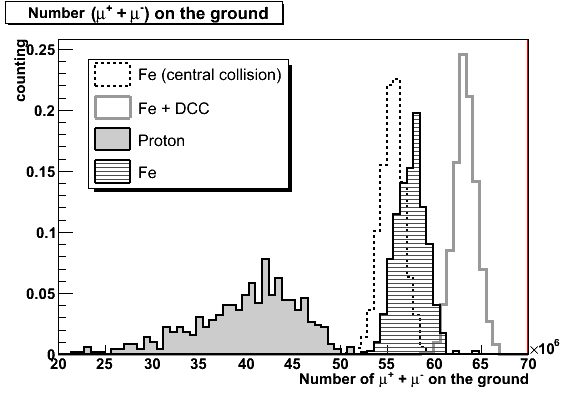}}
		\caption{Normalized distributions of the number of $\mu^{+}$ + $\mu^{-}$ on the ground 
		for vertical showers initiated by Fe (central collision), Fe+DCC, Fe and proton with $E=10^{19}$ eV.}
\label{mu_dist}
	\end{center}
\end{figure}
\begin{figure}[!h]
	\begin{center}
		\scalebox{0.9}{\includegraphics[scale=0.28]{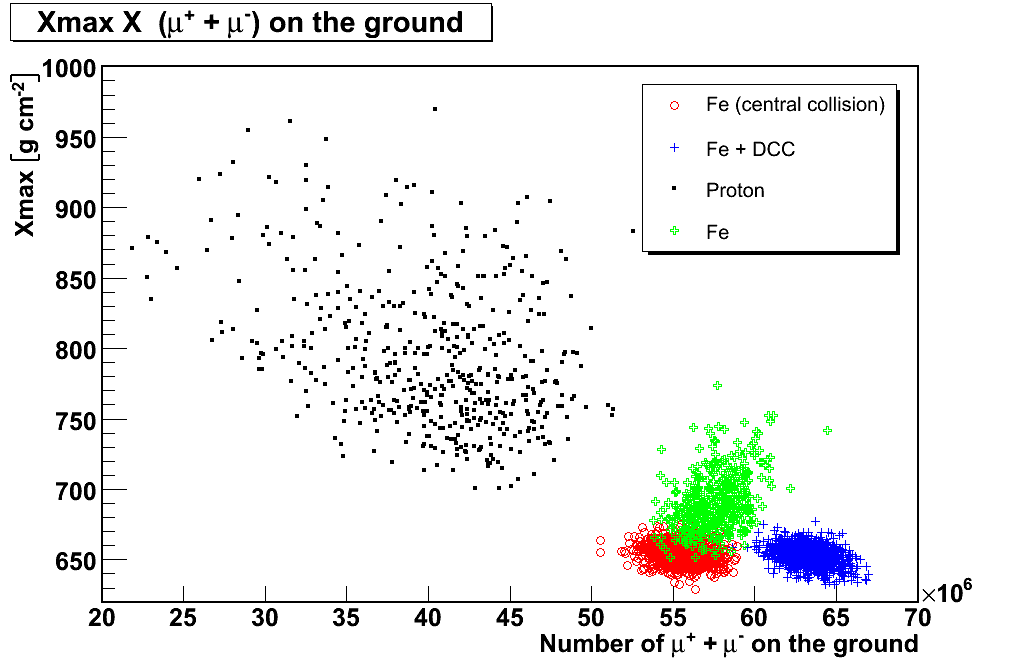}}
		\caption{Depth of the maximum  versus number of $\mu^{+}$ + $\mu^{-}$ on the ground for vertical showers 
initiated by Fe (central collision), Fe+DCC, Fe and proton with $E=10^{19}$ eV.}
\label{scatterplot}
	\end{center}
\end{figure}

The muon number distributions on the ground for the same showers described above are shown in Fig. 
\ref{mu_dist}. As expected, increasing the number of initial charged pions will lead to a corresponding 
larger density of muons on the ground as the products of the charged pions decay.

\begin{figure*}
\center{
\includegraphics[totalheight=2.in, angle=0]{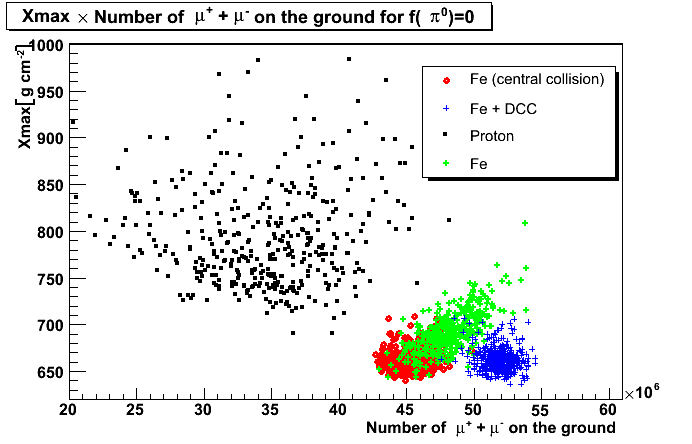}
\includegraphics[totalheight=2.in, angle=0]{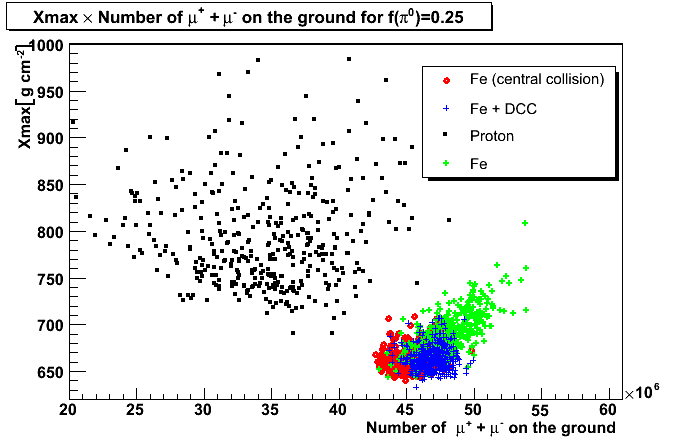}}
\center{
\includegraphics[totalheight=2.in, angle=0]{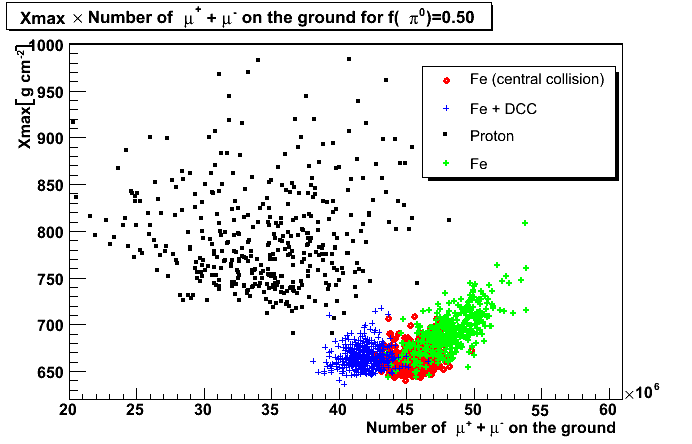}
\includegraphics[totalheight=2.in, angle=0]{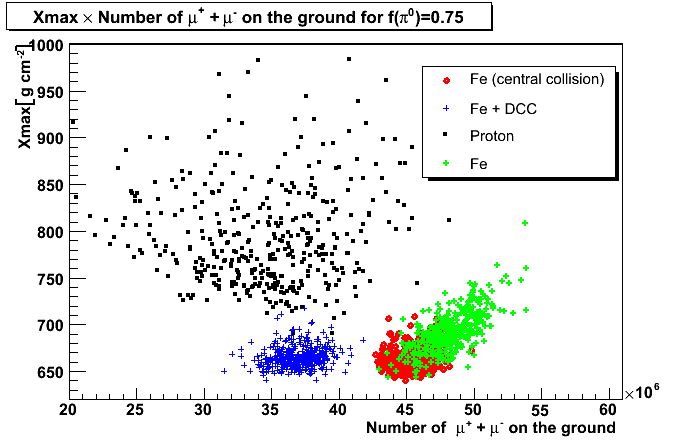}}
\center{
\includegraphics[totalheight=2.in, angle=0]{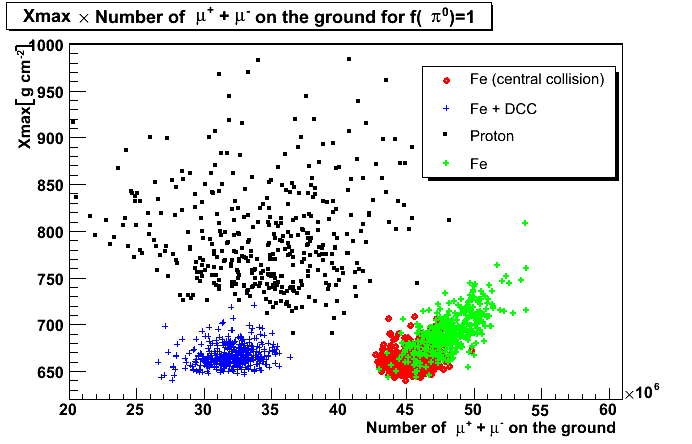}
\includegraphics[totalheight=2.in, angle=0]{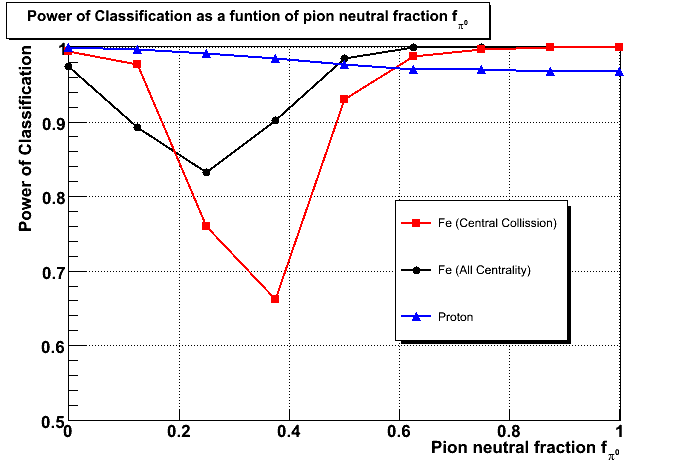}}
\caption{Depth of the maximum  versus number of $\mu^{+}$ + $\mu^{-}$ on the ground for showers with 
$\theta=38$ degrees initiated by Fe (central collision), Fe+DCC, Fe and proton with $E=10^{19}$ eV 
for different $\pi^{0}$ fractions: $f$=0.0 (top left), 0.25 (top right), 0.50 (middle left), 0.75 
(middle right) and 1.0 (bottom left). Bottom right: merit factor (Eq. \ref{meriteq}) as a function 
of $f$.}
\label{merit}
\end{figure*}

As one combines both pieces of information ($X_{max}$ and $N_{\mu}$), one finds 
a good separation between the Fe-DCC 
sample and all the other populations, as can be seen from Fig. \ref{scatterplot}. When collisions 
with all the centralities are allowed, iron showers exhibit a positive correlation between $X_{max}$ 
and $N_{\mu}$. The correlation is such that showers reaching maximum development at increasingly larger 
slant depths produce, in turn, a higher muon density at ground level, due to the decrease in the 
attenuation in the atmosphere from $X_{max}$ to the ground.

Whereas for Fe-Central showers one sees no correlation between $X_{max}$ and $N_{\mu}$, for the case 
of Fe-DCC showers, instead, a small anti-correlation of about 
$1.96$ g/cm$^{2}$/ $10^{6}$ muons is present\footnote{A fit to the binned $X_{max} \times N_{\mu}$ 
distribution provides 
$(1.96 \pm 0.13) \times 10^{-6}$ g/cm$^{2}$/muon.},
as can be 
seen from Fig. \ref{scatterplot}. Such an inverted correlation can be understood remembering the 
additional correlation between $X_{max}$ and the multiplicity in the first interaction, with low 
$X_{max}$ events corresponding, in average, to high multiplicity. As we convert neutral pions to 
charged ones, high multiplicity (small $X_{max}$) events show a larger muon density on the ground 
as compared to low multiplicity ones (large $X_{max}$).

Since the number of muons on the ground depend both on the $\pi^{0}$ fraction $f$ and on the zenith 
angle, due to atmospheric attenuation, we decided to perform a more systematic study by varying those 
parameters. Applying a Linear Discriminant Analysis (LDA), similar to what has been done in 
\cite{Catalani:2007ze} we have determined, as a function of $f$, a power of discrimination defined as 
$1-\beta$, with the classification error $\beta$, in the 2 populations case ($A$ and $B$) given by:
\begin{equation} 
\beta =  \frac{N_{AB}  + N_{BA}}{N_A  +  N_B}
\label{meriteq}
\end{equation}
where $N_{AB}$ ($N_{BA}$) are the number of events from population $A$ ($B$) misclassified as $B$ 
($A$) and $N_{A}$ ($N_{B}$) is total number of events in population $A$ ($B$). As training samples 
we have used half of the simulated showers, with the other half being used for the determination of 
the power of discrimination.

Fig. \ref{merit} (top, middle and bottom left) shows again the two-dimensional 
scatter plots in the $X_{max} \times N_{\mu}$ 
plane for the same 4 populations analyzed above and 5 values of $\pi^{0}$ fraction 
$f=0, 0.25, 0.5, 0.75, 1$ and $\theta=38$ degrees. 
The merit factor as a function of $f$ is shown at the bottom right plot 
of Fig. \ref{merit}. One sees that for any value of $f$, proton showers can always be reasonably 
separated from Fe-DCC, with merit factors above 97\% in the whole interval of $f$. For Fe showers, 
of course, as we approach the $f=1/3$ fraction, the separation power gets weaker. Fe-Central showers 
are easier to separate from the Fe (all centralities) signals only at small values of $f$, 
where the excess of muons 
at ground is large enough to segregate the Fe-DCC population. Fe-DCCs showers in which the first interaction 
produces a large amount of $\pi^{0}$ are very poor in muons and would also be easily recognized. However, 
even if DCCs exist in nature, such large-$f$ events are very unlikely to happen taken into account the 
distribution of Eq. (\ref{sqrt}).

A more robust estimation of the power of discrimination would have to incorporate a priori information 
on the population frequencies for proton and Fe, as well as the DCC occurrence 
frequency. Nonetheless, since 
neither the cosmic rays chemical composition at the highest energies ($E\gtrsim 10^{17}$ eV) nor the full 
dynamics of DCC formation are known, flat priors on these variables are the best one can do so far.

\section{Conclusions}

The recent increase in quality and statistics of UHECR data brings the possibility to look for 
exotic phenomena as well as rare phenomena within the Standard Model. Among the latter, 
DCCs have been predicted about twenty years ago, yet never detected due to their elusive 
character. 

In this paper, using air shower simulations, we searched for fingerprints of DCCs formed in 
ultra-high-energy ($E\sim 10^{19}~$eV) central iron collisions in the upper atmosphere. 
In particular, we studied the influence of the DCC formation on the observables $X_{max}$ 
and $N_{\mu}$, via the asymmetry in the neutral-to-charged pion ratio in the primary collision. 
For comparison, we considered also regular air showers generated by protons and iron.

Since DCCs are expected to be formed in central collisions, which lead to smaller than 
average values of $X_{max}$ (a difference of about $40~$g/cm$^{2}$ as compared 
to iron, for instance), one should concentrate searches within this region of shower depth. 
For the same reason, if DCC events are present they should yield small fluctuations in $X_{max}$. 
However, based only on the behavior of this observable one can not distinguish between 
an iron shower and a central collision and one produced in the presence of a DCC.

The formation of DCCs is associated with large event-by-event fluctuations in the ratio of 
neutral to charged pions, $f$. In particular, for $f$ large or small as compared to $1/3$, 
one should expect a sizable change in the muon density on the ground, especially for 
vertical showers. This fact was noticed in the preliminary study of Ref. \cite{Viana:2007zz}. 
In the extreme case of $f=0$, where DCC events lead to muon-rich showers, we showed 
that this signature distinguishes between the cases of iron (even for central collisions) and 
a DCC event. For large $f$, one can also separate these two cases due to the large depletion 
of muons on the ground. Even in this case, the signature is not contaminated by proton 
events.

For vertical showers, there is a clear anti-correlation between $X_{max}$ and $N_{\mu}$. 
This comes about since there is a correlation between $X_{max}$ and the first interaction 
multiplicity, as was discussed in the previous section. This behavior is not expected in a 
regular iron shower. In fact, for iron, due to atmospheric attenuation, one would expect a 
positive correlation. 

Even though the analysis presented here is very simplified, it has the advantage of 
providing a setup that is totally under control for simulations, and we believe it contains 
the essential ingredients of the phenomena considered. Nevertheless, a more realistic 
study, that should contain a description of the dynamics of DCC formation, especially 
its dependence on energy, is certainly necessary.

\section{Acknowledgments}
We thank F. Catalani and J. Takahashi for useful discussions. We also thank V. de Souza 
for computer time at the USP-S\~ao Carlos cluster [FAPESP, grant 2008/04259-0].
This work was partially supported by by CAPES, CNPq, FAPERJ, FAPESP and FUJB/UFRJ.


\end{document}